\documentclass[useAMS,usenatbib,usegraphicx]{mn2e} 
\bibpunct[, ]{}{}{;}{a}{,}{,} 
\usepackage{graphicx} 
\graphicspath{{plots/}} 
 
\newcommand{\be}{\begin{equation}} 
\newcommand{\ee}{\end{equation}} 
 
\newcommand{\msun}{\mbox{M$_{\sun}$}}

\newcommand{\pder}[2]{\frac{\partial #1}{\partial #2}} 
\newcommand{\pdert}[1]{\pder{#1}{t}} 
\newcommand{\bl}[1]{\mbox{\boldmath$ #1 $}}

\begin{document} 
 
\title[Boltzmann moment equation approach for stellar dynamics]{Boltzmann moment equation 
approach for the numerical study of anisotropic stellar disks} 
\author[E. I. Vorobyov and Ch. Theis]{E. I. Vorobyov$^{1}$\thanks{E-mail: 
vorobyov@astro.uwo.ca (EIV); theis@astro.univie.ac.at (ChT)} and Ch. 
Theis$^{2}$ \\ 
$^{1}$CITA National Fellow, Department of Physics and Astronomy, University of Western Ontario, 
London, Ontario, N6A 3K7, Canada. \\ 
$^{2}$Institut f\"ur Astronomie, Universit\"at Wien, 
              T\"urkenschanzstr. 17, 1180 Wien, Austria.}

\date{to appear MNRAS} 
 
\maketitle 
 
\label{firstpage}

\begin{abstract} 
  
 We present the Boltzmann moment equation approach for the dynamics of stars (BEADS-2D), 
which is a finite-difference Eulerian numerical code designed for the modelling of 
{\it anisotropic and non-axisymmetric} flat stellar disks. The BEADS-2D code 
solves the Boltzmann moment equations up to second order in the thin-disk approximation. 
This allows us to obtain the anisotropy of the velocity ellipsoid 
and the vertex deviation in the plane of the disk.

We study the time-dependent evolution of exponential stellar disks 
in the linear regime and {\it beyond}. The disks are initially characterized 
by different values of the 
Toomre parameter $Q_{\rm s}$ and are embedded in a dark matter halo, 
yielding a rotation curve composed of a rigid central part and a 
flat outer region. Starting from a near equilibrium state, several 
unstable modes develop in the disk. 
In the early linear phase, the very centre and the large scales are characterized 
by growing one-armed and bisymmetric positive density perturbations, respectively.  
This is in agreement with expectations from the swing 
amplification mechanism of short-wavelength trailing disturbances, 
propagating through the disk centre. In the late linear phase, the overall 
appearance is dominated by a two-armed spiral structure localized within 
the outer Lindblad resonance (OLR).  During the non-linear evolutionary phase, 
radial mass redistribution due to the gravitational torques of spiral arms 
produces an outflow of mass, which forms a ring at the OLR, and an inflow 
of mass, which forms a transient central bar. 
This process of mass redistribution is self-regulatory and it terminates 
when spiral arms diminish due to a 
shortage of matter. Finally, a compact central disk and a diffuse ring at the OLR 
are formed. An increase in $Q_{\rm s}$ stabilizes the 
disks at $Q_{\rm s}\approx 3.15$, in agreement with the theoretical predictions. 
 
Considerable vertex deviations are found in regions with strongly 
perturbed mass distributions, i.e.\ near the spiral arms.  The 
vertex deviations are especially large at the convex edge of the 
spiral arms, whereas they are small at the concave edge.  The mean 
vertex deviations correlate well with the global Fourier amplitudes, 
reaching mean values of about 12$^\circ$  in the saturation stage. 
Local values of the vertex deviation can reach up to almost 90$^\circ$. 
Near the convex edge of the spiral arms, the ratio of radial to azimuthal components 
of the velocity ellipsoid can deviate considerably from 
the values predicted from the epicycle approximation. 
 
\end{abstract} 
 

\section{Introduction} 
 
   The majority of normal disk galaxies are characterized by 
non-axisymmetric structures like spirals or bars. 
These structural elements have been widely discussed in the literature 
as a result of gravitational instabilities which are connected to 
growing density waves or global instabilities of disks 
(e.g.\ Binney \& Tremaine 1987, afterwards BT87). 
 
A first insight into the properties of galactic discs was provided by linear 
stability analysis. Structure growth in a stellar disk within the linear regime was 
studied via the linearised collisionless Boltzmann equation 
by Toomre (\cite{toomre64}), Zang (\cite{Zang}),  Pichon \& Cannon (\cite{Pichon}), 
Evans \& Read (\cite{Evans}), Jalali \& Hunter (\cite{Jalali}), and others. 
Toomre (\cite{toomre64}) has deduced a simple 
stability criterion from the Jeans equations in case of stellar disks: 
a stellar disk is stable against axisymmetric perturbations if 
$Q_{\rm s} \equiv \sigma_{rr} \kappa / (3.36 G \Sigma)$ exceeds 
a critical value $Q_\mathrm{s,crit}=1$ ($\sigma_{rr}$ is the radial velocity 
dispersion, $\kappa$ is the epicycle frequency, $G$ is the gravitational constant, and $\Sigma$ 
is the surface density of the disk). 
In case of non-axisymmetric perturbations the 
analysis becomes technically more difficult, but the Toomre parameter 
turned out to still be a good indicator for stability, if one increases its 
critical value (Bertin et al.\ \cite{bertin89}, 
Polyachenko et al.\  \cite{polyachenko97b}).

   A disadvantage of linear stability analysis is its restriction 
to small perturbations, both in amplitude and wavelength. 
Once the perturbations exceed a critical level or when the initial perturbations 
are already large, numerical simulations are necessary. 
Hydrodynamical simulations revealed three qualitatively 
distinct phases for the growth of unstable disks 
(Laughlin et al.\ \cite{laughlin97}). Starting from small amplitudes, 
perturbations grow first linearly. When they reach a critical level, 
mode-mode coupling becomes important and the non-linear evolution starts. 
The self-interaction of the dominant mode might 
result in a rearrangement of the radial mass distribution ($m=0$ mode). 
The disk itself might become violently unstable, which is indicated by the 
perturbations reaching a ``macroscopic'' level. Finally, a quasi-equilibrium saturation stage can be reached 
when the growth of the instabilities is stopped due to shocks or 
feedback mechanisms heating up the disk dynamically. Otherwise, fragmentation 
continues until the disk is destroyed or transformed.

   Though N-body simulations have been proven to be extremely useful for 
many studies of galactic dynamics, they are less suitable 
for studying the growth of very weak perturbations. The main reason is 
that N-body calculations of galaxies are usually performed with less 
particles than the actual number of stars in these systems. This causes an inherent artificial 
particle noise which cannot be neglected for typical particle numbers. 
    
   Alternatively, numerical hydrodynamics simulations became a primary tool 
for the analysis of growing spiral instabilities (e.g. Korchagin et al.\ 
\cite{korchagin00}, Orlova et al.\ 
\cite{orlova02}). First, they allow for a direct comparison with the results of a semi-analytic 
linear stability analysis, including predictions for the growth rates 
(Laughlin et al. \cite{laughlin97,laughlin98}, Korchagin et al.\ \cite{korchagin00}), 
because the same set of basic equations is applied. 
Second, the standard grid methods used to solve the hydrodynamical 
equations allow to start with very small perturbations (basically 
limited by machine accuracy). Their evolution can be followed over many 
$e$-folding times until they reach the non-linear regime and beyond. 
 
   A major drawback of purely hydrodynamical simulations 
is that the dominant disk component is the stellar disk. 
The latter is characterized by an {\it anisotropic 
velocity dispersion}, whereas the hydrodynamical equations deal with 
an {\it isotropic pressure}. Secondly, one has to assume an equation of state 
connecting pressure, (surface) density and temperature (velocity dispersion). 
For instance, if one assumes a polytropic-like equation of state, i.e.\ 
$P = C \, \Sigma^\gamma$, 
the set of hydrodynamic equations is closed with the Euler equation 
describing the momentum transport 
(for an example of a stability analysis 
based on such assumptions see e.g.\ Aoki et al.\ \cite{aoki79}). 
Since the temperature does not show 
up in the polytropic equation of state, no energy transport equation is required. 
Obviously, this is a convenient simplification which, however, might 
not be generally applicable. Moreover, the profile of the Toomre parameter 
is completely fixed by the surface density distribution. 
 
  A more complete ansatz for describing stellar disks 
is the collisionless Boltzmann equation (BT87) 
\be 
   \label{eq_boltzmann} 
   \pdert{f} + \bl{v} \cdot \pder{f}{\bl{r}} + 
       \dot{\bl{v}} \cdot \pder{f}{\bl{v}} = 0, 
\ee 
where $f(\bl{r},\bl{v},t)$ is the distribution function. The term 
$\dot{\bl{v}}$ includes accelerations like gravitational forces, 
pressure terms, etc.  Since a direct solution is already numerically 
very difficult, the problem is often reduced by taking the velocity 
moments of equation~(\ref{eq_boltzmann}).  By this, one gets an 
infinite series of moment equations where the equations for moment $i$ 
include moments of the order $i+1$.  The knowledge of all velocity 
moments is equal to the knowledge of the distribution function itself. 
Practically, one has to terminate the set of equations by a closure 
condition. Setting all moments of third order to zero (zero-heat-flux 
condition) results in the Jeans equations which are well applicable to 
stellar systems with negligible two-body relaxation.  In that case, 
the second order moment equations which yield information on the 
velocity ellipsoid are included and stellar anisotropy can be 
considered.  An example are the one- and two-dimensional 
chemo-dynamical models by Theis et al.\ (\cite{theis92}) and Samland et 
al.\ (\cite{samland97}).  For the special case of galactic disks, Amendt 
\& Cuddeford (\cite{amendt91}) and Cuddeford \& Amendt 
(\cite{cuddeford91}) studied the higher order moments in detail. They 
found that for reasonable constraints to the distribution function 
the third order terms vanish to leading order in the plane 
of the disk. This corroborates the assumption of a zero-heat-flux. 
Extensions of the Boltzmann moment equations including collisional 
processes have also been used (e.g.\ Larson \cite{larson70}, Louis 
\cite{louis90} or Giersz \& Spurzem \cite{giersz94} for the evolution 
of star clusters).

 
  In this paper we present the Boltzmann equation approach for the dynamics 
of stars (BEADS-2D), which is a numerical code that solves the Boltzmann moment equations 
up to second order using the methods of finite-differences. 
The BEADS-2D code is applied to study the time-dependent evolution of flat, 
non-isotropic stellar disks in the linear regime and {\it beyond}. 
In contrast to previous numerical studies, we include the {\it non-axisymmetric} 
velocity dispersion terms into the Boltzmann moment equations. 
Therefore, our approach allows us to study the evolution 
of the anisotropy and the vertex deviation of the stellar component. 
We compare the stability properties of non-isotropic stellar disks with 
the predictions of linear stability analysis for non-isotropic 
disks by Polyachenko et al. (\cite{polyachenko97b}). 
We provide a physical interpretation for growing instabilities 
in our model stellar disks. 
 
 
In the following section we describe the Boltzmann moment equation approach. 
The basic principles of the BEADS-2D code are discussed in \S~\ref{sect_numericalmodelling}. 
The results of numerical modelling  are presented in \S~\ref{sect_results}. 
Finally, conclusions 
and a summary are given in \S~\ref{sect_summary}. 
 
 
\section{Stellar systems: the moment equation approach} 
\label{sect_mathematicalmodel} 
 
\subsection{The 3D-Boltzmann moment equations} 
 
Stars are collisionless objects that move on orbits determined 
by the large-scale gravitational potential $\Phi$. An exact description of such a 
system requires the solution of the collisionless Boltzmann equation (\ref{eq_boltzmann}) 
for the distribution function of stars $f({\bl r},{\bl v},t)$ in the phase space 
($\bl{r},\bl{v}$). 
The distribution function of stars $f\equiv f({\bl r},{\bl v},t)$ 
contains a fundamental description of the stellar system. 
Its lower order moments like the density $\rho=\int f \,d^3{\bl v}$, 
the mean velocity $\bl{u}=\rho^{-1} \int f \,{\bl v}\, d^3{\bl v}$, 
or the velocity-dispersion tensor 
$\sigma^2_{ij}=\rho^{-1}\int f\, (v_i-u_i)(v_j-u_j)\,d^3{\bl v}$ 
of stars can be deduced -- at least partly -- from observations. 
The corresponding equations for $\rho$, $\rho {\bf u}$, and $\rho \sigma^{2}_{ij}$ 
can be obtained by taking moments of the collisionless Boltzmann equation 
(\ref{eq_boltzmann}). If one closes the system by assuming the zero-heat-flux 
approximation $Q_{ijk}=\rho^{-1}\int f\,(v_i-u_i)(v_j-u_j)(v_k-u_k)\,d^3{\bl v}=0$, 
a set of ten partial differential equations (the velocity dispersion tensor is 
symmetric by definition) can be derived (see Appendix~A). 
 
The three off-diagonal elements of the velocity-dispersion tensor can be regarded 
as a measure of alignment of the principal axes of the velocity ellipsoid 
with respect to the coordinate system. We note that these terms may become 
both positive or negative (although they have a square mark), as 
follows directly from the definition of the velocity-dispersion tensor.

In our models, the gravitational potential $\Phi$ is  composed of the sum of an 
external contribution $\Phi_{\rm ext}(\bl{r},t)$ and the 
self-gravity $\Phi_{\rm sys}(\bl{r},t)$ of the disk mass distribution $\rho(\bl r)$ 
obeying the moment equations. Its potential can be derived from 
the Poisson equation 
\begin{equation} 
   \Delta \Phi_{\rm sys} = 4 \pi G \rho. 
\end{equation} 
 The external term describes any known 
(eventually time-dependent) potential like a dark matter halo, 
whereas the ``system'' corresponds to the live disk. 
 
 
\subsection{The 2D-Boltzmann moment equations for flat disks} 
 
A substantial fraction of stars in spiral galaxies are concentrated 
in the disk. Hence to a 
first approximation, stellar disks may be regarded as having zero 
thickness. In the thin-disk approximation, all motions are localized 
within the ($r,\phi$) plane. 
The moment equations in the thin-disk approximation 
can be obtained from equations~(\ref{eq_3dkin_rho})-(\ref{eq3d_sigma_rz}) by assuming 
a vanishing $z$-gradient of all physical variables and setting $u_z=0$ 
(negligible vertical motion), $\sigma^2_{zr}=0$, and $\sigma^2_{\phi z}=0$. 
The stellar volume density $\rho$ is vertically integrated to yield the 
surface density $\Sigma$. 
The assumptions of $\sigma^2_{zr}=0$, and $\sigma^2_{z\phi}=0$ 
are justified on observational grounds (Binney \& Merrifield \cite{binney98}).   
The vertical velocity dispersion $\sigma_{zz}$ is obtained by assuming 
a constant ratio of the vertical to the radial (azimuthal) velocity dispersions. 
Below, we provide the moment equations in the thin-disk approximation written 
in polar coordinates. For the convenience of coding, the advection 
terms are set in brackets.   
\\ 
 
\noindent 
{\bf Continuity equation:} 
   \begin{equation} 
     {\partial \Sigma \over \partial t} + 
      \left[{1 \over r} {\partial \over \partial r} 
         (r\cdot \Sigma \cdot u_r) + {1 \over r} {\partial \over \partial \phi} 
         (\Sigma \cdot u_{\phi}) 
      \right] = 0. 
   \label{eq_2dkin_rho} 
\end{equation} 
 
\noindent 
{\bf Momentum equations:} 
\begin{eqnarray} 
      {\partial \over \partial t}(\Sigma u_{r})  & + & 
       \left[ 
          {1 \over r} {\partial \over \partial r} 
                (r \cdot \Sigma u_{r} \cdot u_{r}) 
        + {1 \over r} {\partial \over \partial \phi} 
                (\Sigma u_{r} \cdot u_{\phi}) 
       \right] \nonumber \\   
   & - & 
       {\Sigma u_{\phi}^2 \over r}  +   
       {1\over r} {\partial \over \partial r} (r \Sigma \sigma_{rr}^2) + 
       {1 \over r}{\partial \over \partial \phi} (\Sigma \sigma_{r \phi}^2) 
    \nonumber \\ 
   & - & 
       {\Sigma \sigma_{\phi\phi}^2 \over r} + 
        \Sigma {\partial \Phi \over \partial r}  = 0, 
    \label{eq_2dkin_ur} 
\end{eqnarray} 
 
\begin{eqnarray} 
    {\partial \over \partial t} (\Sigma r u_{\phi }) 
    & + & \left[ 
              {1\over r}  {\partial \over \partial r} 
                    (r \cdot \Sigma r u_{\phi} \cdot u_{r}) 
            + {1\over r } {\partial \over \partial \phi} 
                    (r \Sigma u_{\phi} \cdot u_{\phi})   
          \right] 
     \nonumber \\ 
    & + & r \left\{ 
                {\partial \over \partial r} (\Sigma \sigma_{\phi r}^2) 
              + {1\over r} {\partial \over \partial \phi} (\Sigma \sigma_{\phi \phi}^2) 
              + {2\over r} \Sigma \sigma_{r \phi}^2 
            \right\} \nonumber \\ 
    & + & \Sigma {\partial \Phi \over \partial \phi} = 0. 
    \label{eq_2dkin_uphi} 
\end{eqnarray} 
 
\noindent 
{\bf Velocity dispersion equations:} 
\begin{eqnarray} 
     {\partial \over \partial t}(\Sigma \sigma_{rr}^2) 
     &+&  \left[ 
             {1 \over r}  {\partial \over \partial r} 
                          (r\cdot \Sigma \sigma_{rr}^2 \cdot u_{r}) 
           + {1 \over r} {\partial \over \partial \phi} 
                          (\Sigma \sigma_{rr}^2 \cdot u_{\phi}) 
          \right]  \nonumber \\ 
      &+& 2 \Sigma \sigma_{r\phi}^2 
             \left\{ {1\over r} {\partial u_{r} \over \partial \phi}   
                   - {2 u_{\phi}\over r} 
             \right\}  \nonumber \\ 
      &+& 2 \Sigma \sigma_{rr}^2 {\partial u_{r} \over \partial r} = 0, 
   \label{eq_2dkin_sigmarr} 
\end{eqnarray} 
 
\begin{eqnarray} 
       {\partial \over \partial t} (\Sigma \sigma_{\phi \phi}^2)   
    & + & 
       \left[ 
         {1\over r} {\partial \over \partial r} 
              (r \cdot \Sigma \sigma_{\phi \phi}^2 \cdot u_{r}) 
       + {1 \over r} {\partial \over \partial \phi} 
              (\Sigma \sigma_{\phi \phi}^2 \cdot u_{\phi}) 
       \right] \nonumber \\ 
    & + & 2 \Sigma \sigma_{\phi\phi}^2 
          \left\{ {u_{r} \over r} 
                + {1\over r} {\partial u_{\phi} \over \partial \phi } 
          \right\} 
     \nonumber \\ 
    & + & 2 \Sigma \sigma_{r\phi}^2 
          \left\{ {u_{\phi} \over r} 
                + {\partial u_{\phi} \over \partial r } 
          \right\} = 0, 
   \label{eq_2dkin_sigmaphiphi} 
\end{eqnarray} 
 
\begin{eqnarray} 
        {\partial \over \partial t} (\Sigma \sigma_{r \phi}^2) 
     & + & 
       \left[ 
          {1\over r} {\partial \over \partial r} 
               (r \cdot \Sigma \sigma_{r \phi}^2 \cdot u_{r}) 
           + {1 \over r} {\partial \over \partial \phi} 
               (\Sigma \sigma_{r \phi}^2 \cdot u_{\phi}) 
       \right] 
       \nonumber \\ 
     & + & \Sigma \sigma_{r\phi}^2 \, 
        \left\{ {1\over r} {\partial  \over \partial r} (r u_{r}) 
              + {1\over r}{\partial u_{\phi} \over \partial \phi} 
        \right\} 
     \nonumber \\ 
     & + & \Sigma \sigma_{rr}^2 
        \left\{ {u_{\phi} \over r} 
              + {\partial u_{\phi} \over \partial r} 
        \right\} 
        \nonumber \\ 
      & + & \Sigma \sigma_{\phi\phi}^2 
        \left\{ {1 \over r}{\partial u_{r} \over \partial \phi} 
              - {2 u_{\phi} \over r} 
        \right\} = 0. 
     \label{eq_2dkin_sigmarphi} 
\end{eqnarray} 
According to our experience, equations~(\ref{eq_2dkin_rho})-(\ref{eq_2dkin_sigmarphi}) are expressed 
in the computationally most stable form. 
For convenience we will use the name {\it kinetic equations} or 
{\it kinetic models} when we solve the second order Boltzmann moment 
equations.

\subsection{The gravitational potential} 
\label{sect_potential} 
 
In our models the gravitational potential consists of two parts, a live disk 
and a static halo component. 
The gravitational potential $\Phi_{\rm disk}$ 
of the flat disk can be calculated by (BT87, Sect.\ 2.8) 
\begin{eqnarray} 
  \Phi_{\rm disk}(r,\phi) & = & - G \int_0^\infty r^\prime dr^\prime 
                     \nonumber \\ 
      & &       \times \int_0^{2\pi} 
               \frac{\Sigma(r^\prime,\phi^\prime) d\phi^\prime} 
                    {\sqrt{{r^\prime}^2 + r^2 - 2 r r^\prime 
                       \cos(\phi^\prime - \phi) }}  \, . 
\end{eqnarray} 
This sum is calculated using a FFT technique which applies the 2D Fourier 
convolution theorem for polar coordinates. 
 
The halo properties are fixed by the rotation curve parametrized by 
\begin{equation} 
   v_{\rm c} = v_\infty \cdot \left( \frac{r}{r_{\rm flat}} \right) 
                 \cdot \frac{1} 
                   {\displaystyle 
                    \left[ 1 + \left( \frac{r}{r_{\rm flat}} \right)^{n_v} 
                    \right]^{\displaystyle \frac{1}{n_v}}} \,\,. 
   \label{vcirc} 
\end{equation} 
The transition radius between an inner region of rigid rotation 
and a flat rotation in the outer part is given by $r_{\rm flat}$ which 
we set to 3 kpc. The smoothness of the transition is controlled by the parameter 
$n_v$, set to 2. The velocity at infinity, $v_\infty$, was set to 208 km~s$^{-1}$. 
The solid line in Fig.~\ref{fig1} shows the rotation curve we used. 
The corresponding halo potential $\Phi_{\rm halo}$ is then derived from 
equation~(\ref{eq_2dkin_ur}), where $\Phi$ is substituted with $\Phi_{\rm halo}+\Phi_{\rm disk}$ 
and $u_{r}$ is set to zero.


\section{Numerical modelling} 
\label{sect_numericalmodelling}

\subsection{The BEADS-2D code} 
\label{code2D} 
The 2D-equations for thin disks are discretized on an Eulerian grid 
with a logarithmic grid spacing in the $r$-direction and uniform grid spacing 
in the $\phi$-direction. The different terms 
are taken into account by applying an operator splitting technique 
similar to the ZEUS program (Stone \& Norman \cite{stone92}). Advection 
is performed by the second-order van-Leer scheme. The time step is determined 
according to the usual Courant-Friedrichs-Levy criterion 
modified to take into account the velocity dispersion. Namely, 
we diagonalize the velocity dispersion tensor at each time step and use 
the diagonal components in the time-step limiter. 
For our simulations we use reflecting boundary conditions in the radial direction 
(and periodic boundary conditions in the azimuthal direction).   
Our simulations are done on a grid with $256\times256$ cells 
covering the radial range from 0.2 kpc to 30-45 kpc (depending on the 
model). The gravitational force of 
a thin disk near its inner boundary at $r=0.2$~kpc is directed radially 
outward, which is an artificial effect due to 
the lack of material within the inner boundary (note that such a problem 
would not exist in the case of spherical symmetry). 
In order to reduce the magnitude of this spurious 
outward gravity force, we introduce an inner circular 
disk of constant density and outer radius $r_{\rm out}=0.2$~kpc. This inner 
disk merely serves as a central gravity source and its gravitational potential 
$\Phi_{\rm id}$ can be computed by decomposing the inner disk into 
a series of concentric circular rings of constant density (see 
Vorobyov \& Basu \cite{VB} for details). A more accurate method that is 
used in this paper involves an expansion of $\Phi_{\rm id}$ into Legendre 
polynomials in the plane of the disk. The details of this method 
will be given in a future paper. 
The total gravitational potential $\Phi=\Phi_{\rm disk}+\Phi_{\rm halo}+\Phi_{\rm id}$ 
is then used in the momentum equations. 
 
 
\subsection{Artificial viscosity} 
The artificial viscosity, much like kinematic viscosity in a real fluid, 
is often used to smooth discontinuities where the finite-difference equations 
break down. The 2nd order Boltzmann moment equations have to be modified to account for the viscous 
stresses and dissipation due to the artificial viscosity. 
 
 
The implementation of the artificial viscosity in the kinetic model 
requires the use of a tensor formalism. The artificial viscosity stress 
tensor $\bl Q$ can be written by analogy to the general form of the molecular 
viscosity stress tensor 
\begin{equation} 
{\bl Q}=2 \mu_{\rm v} [{\bl \nabla \bl u} - {1 \over 3} ({\bl\nabla \cdot \bl u}) \,{\bl 
e}], 
\end{equation} 
where $\bl \nabla \bl u$ is the symmetrized velocity gradient tensor, $\bl 
\nabla \cdot \bl u$ is the velocity divergence, $\bl 
e$ is the unit tensor, and $\mu_{\rm v}$ is coefficient of artificial viscosity 
defined as 
\[ \mu_{\rm v} = \left\{ \begin{array}{ll} 
   L^2 \Sigma\, {\bl \nabla \cdot \bl u} & \,\, \mbox{if $\bl \nabla \cdot \bl 
   u<0$} \\ 
   0 & \,\, \mbox{otherwise},   \end{array} 
   \right. \] 
where $L$ is a constant with dimensions of length, chosen to 
be one zone width. 
The components of the artificial viscosity stress tensor $\bl Q$ in 
cylindrical coordinates are given in Appendix B.   
The off-diagonal components of $\bl 
Q$ are often set to zero to ensure that the artificial viscosity smoothes only compressible 
shock fronts and leaves large gradients of shear unchanged. 
The corresponding 
viscous stress and dissipation terms are introduced into the kinetic 
equations~(\ref{eq_2dkin_ur})-(\ref{eq_2dkin_sigmarphi}) by 
defining a generalized  stress tensor $P_{ij}=\Sigma \sigma_{ 
ij}+Q_{ij}$, where the artificial viscosity stress tensor $Q_{ij}$ is analogous 
to the viscous stress tensor $-\pi_{ij}$ in gas dynamics. 
We note that $Q_{ij}$ has a plus sign because the coefficient of artificial 
viscosity $\mu_{\rm v}$ is negative by definition. The modified momentum equations 
can be obtained from equations~(\ref{eq_2dkin_ur}) and (\ref{eq_2dkin_uphi}) 
by substituting $\Sigma \sigma_{rr}$ with $P_{rr}$ and $\Sigma \sigma_{\phi\phi}$ 
with $P_{\phi\phi}$. The modified velocity dispersion equations 
can be derived by applying the same substitution to the last two terms of 
equations (\ref{eq_2dkin_sigmarr}) and (\ref{eq_2dkin_sigmaphiphi}) and 
to the last three terms in equation~(\ref{eq_2dkin_sigmarphi}).

\subsection{Units}

  The units are chosen to be 
  $10^{10}~\msun$ and 1 kpc for the mass and length scales, respectively. 
  The gravitational constant $G$ was set to 1. This results in a 
  velocity unit of 207.4 km\,s$^{-1}$ and a unit for the 
  angular speed of 207.4 km\,s$^{-1}$\,kpc$^{-1}$. The time unit 
  is 4.7 Myr. These units are used throughout the paper, unless 
  other units are given explicitly. 
 
 
\subsection{Tests and accuracy} 
 
   We performed several tests for our numerical code. Two standard tests -- 
one focusing on the accuracy of the implementation of the advection 
and another one on the conservation of the specific angular momentum -- 
are given in Appendix~C in some detail. Both tests demonstrate the ability of the 
code to handle the advection terms numerically well, so that we do not 
expect spurious structure formation due to advection problems. 

\subsection{Fourier amplitudes} 
 
   In order to visualize the nature and growth of instabilities in 
disks, it is common to use the azimuthal Fourier modes $a_{\rm m}$ at a given 
radius $r$. They are calculated for $m \neq 0$ by 
(e.g.\ Laughlin et al.\ \cite{laughlin98}) 
\begin{equation} 
   a_{\rm m}(r,t) \equiv \frac{1}{2\pi} 
          \left| \int_0^{2\pi} \Sigma(r,\phi,t) e^{im\phi} d\phi \right| \,\, . 
\end{equation} 
An even simpler measure for the global structure of a disk 
are the (radially integrated) {\it global Fourier modes} defined by 
\begin{equation} 
   C_{\rm m}(t) \equiv \frac{2\pi}{M_{\rm disk}} 
     \int_{r_{\rm low}}^{r_{\rm high}} a_{\rm m}(r,t) r dr \,\, . 
\end{equation} 
$M_{\rm disk}$ is the mass of the disk within the radial range 
$[r_{\rm low},r_{\rm high}]$. The global Fourier amplitude $A_m$ is 
the modulus of $C_{\rm m}$, whereas the global growth rate is given 
by the time derivative of the logarithmic Fourier amplitude, i.e.\ 
\begin{equation} 
  \gamma_m \equiv \frac{d (\ln |C_{\rm m}(t)|)}{dt} \,\, . 
\end{equation}

\begin{figure} 
   \resizebox{\hsize}{!}{ 
     \includegraphics[angle=0]{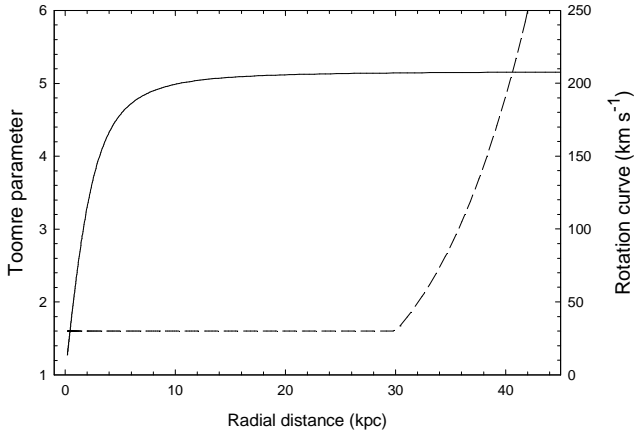}}     
   \caption{Initial profiles of the 
         rotation curve (solid line, identical for all models), 
         and Toomre parameter (dashed line, model~K3) 
          We note that the $Q$-profile just scales 
          in different models.} 
   \label{fig1} 
\end{figure}

\section{Results} 
\label{sect_results} 
 
\begin{table} 
\caption{Model parameters of stellar disks} 
\label{Table2} 
\vskip 0.1cm 
\begin{tabular}{lllc} 
\hline 
Model & $Q_{s}$ & $\sigma_{rr}(r=0)$ & $R_{\rm out}$ (kpc)\\ [2 pt] 
\hline 
K1 & 1.1 & 112.5~km~s$^{-1}$ & 30 \\ 
K2 & 1.3 & 133~km~s$^{-1}$ & 30\\ 
K3 &1.6& 164~km~s$^{-1}$  & 45 \\ 
K4 &2.5& 250~km~s$^{-1}$ & 45\\ 
K5 & 3.15 & 315~km~s$^{-1}$ & 45 \\ 
\hline 
\end{tabular} 
 
\end{table}

\subsection{Stability properties of non-isotropic (stellar) disks} 
\label{sect_results_constQ} 
 
   The analysis of the stability of a {\it thin} stellar disk to a local 
{\it axisymmetric} perturbation states that the disk is stable 
if (Toomre \cite{toomre64}) 
\begin{equation} 
    Q_{\rm s}\equiv {\sigma_{rr} \kappa \over 3.36 G \Sigma} >1. 
    \label{eq_toomre} 
\end{equation} 
According to Polyachenko et al.\ 
(\cite{polyachenko97b}), the Toomre criterion (\ref{eq_toomre}) should 
be modified to $Q_{\rm s} > 2\pi \beta/3.36$, if local 
{\it non-axisymmetric } perturbations are considered. Here, $\beta$ is a 
function of the rotation curve.  In the most interesting case of a 
flat rotation curve, $\beta=1.69$, so that a thin stellar disk becomes 
stable if $Q_{\rm s}\ga 3.15$.

In this section we numerically study the stability properties of thin stellar disks 
characterized by different values of $Q_{\rm s}$. 
We do not focus on the detailed study of growth rates of unstable modes 
in the linear regime, 
since it would require a detailed comparison with previous analytical linear stability 
analyses. Instead, we simply compare the stability properties of stellar disks 
obtained using the BEADS-2D code with the analytical predictions 
of Polyachenko et al. ({\cite{polyachenko97b}). 
By using the kinetic equations (\ref{eq_2dkin_rho})-({\ref{eq_2dkin_sigmarphi}), we take full 
account of the inherent anisotropic properties of stellar disks. 
We present the results for five models, the parameters of which are shown 
in Table~\ref{Table2}.  The acronym K stand for ``Kinetic models''. 
The outer radii of stellar disks $R_{\rm out}$ were chosen so that to exclude the influence 
of the outer reflecting boundary. 
The surface density distribution and the rotation curve are 
identical for all models, while the initial velocity dispersion distributions 
are different.  The surface density is distributed exponentially 
according to 
\begin{equation} 
   \Sigma(r) = \Sigma_0 e^{-r/r_d}, 
   \label{surfdens} 
\end{equation} 
with a radial scale length $r_d$ of 4 kpc. The central surface density 
$\Sigma_0$ is set to $10^3~\msun~\mathrm{pc}^{-2}$. Thus, the total mass 
of the disk within the 30~kpc radius is approximately $10^{11} \msun$. We note that the 
densities of stellar disks are indeed found to decay exponentially with distance, 
with a characteristic scale length increasing from 2-3~kpc for the early type galaxies 
to 4-5~kpc in the late type galaxies (Freeman \cite{freeman}). 
The initial azimuthal velocity is chosen according to equation~(\ref{eq_2dkin_ur}), 
whereas the radial velocity vanishes initially. The radial component of the velocity dispersion 
is obtained from the relation $\sigma_{rr}=3.36 \,Q_{\rm s}\, G \,\Sigma/\kappa$ 
for a given value of the Toomre parameter $Q_{\rm s}$. We assume that throughout 
most of the disk $Q_{\rm s}$ is constant (and equal to the value given 
in Table~\ref{Table2}) but is steeply increasing with radius at $r>30$~kpc. 
The initial radial distribution of $Q_{\rm s}$ in model~K3 is 
shown in Fig.~\ref{fig1} by the dashed line. The azimuthal 
component of the velocity dispersion $\sigma_{\phi\phi}$ is determined adopting 
the epicycle approximation, in which 
the following relation between $\sigma^2_{\phi\phi}$ and $\sigma^2_{rr}$ holds 
(Binney \& Tremaine \cite{binney87}, p.\ 125): 
\begin{equation} 
   \sigma_{\phi\phi}^2=\sigma_{rr}^2 \cdot {1\over2} \left({r \over u_\phi} 
   {du_\phi \over dr}+1 \right). 
   \label{epic} 
\end{equation} 
The off-diagonal component $\sigma^2_{r\phi}$ is initially 
set to zero. In the beginning of simulations we introduce a small random perturbation 
with a maximum relative amplitude $10^{-5}$ to the initial density distribution. 
 
\begin{figure} 
   \resizebox{\hsize}{!}{ 
     \includegraphics[angle=0]{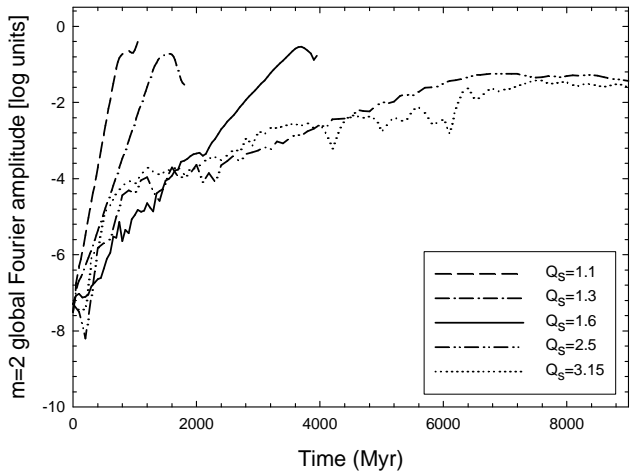} 
   } 
   \caption {The temporal behaviour of the $m=2$ mode in kinetic models with different initial 
   $Q$-parameters: $Q_{\rm s}=1.1$ (K1), 1.3 (K2), 1.6 (K3), 2.5 (K4), and 3.15 (K5).} 
   \label{fig2} 
\end{figure}

Figure~\ref{fig2} shows the temporal behaviour of the dominant $m=2$ mode in models K1-K5 with 
different initial values of the $Q$-parameter. 
It is evident that models with initially larger $Q_{\rm s}$ attain the 
saturation phase at later times than do the models with smaller $Q_{\rm s}$. 
For instance, in model~K2 ($Q_{\rm s}=1.3$) the saturation of 
the dominant mode is reached after 1.4~Gyr, 
whereas in model~K1 ($Q_{\rm s}=1.1$) the dominant mode saturates at an earlier time $t=0.8$~Gyr. 
Model~K4 with the initial $Q_{\rm s}=2.5$ reaches the saturation phase 
only after approximately 7~Gyr.  The global Fourier amplitudes in the saturation phase 
are noticeably larger in models K1--K3 ($C_2(t) \approx 0.1-0.25$) than in models K4 and K5 
($C_2(t) \approx 0.02-0.05$), indicating a stronger instability in colder disks.

 The linear stability analysis of non-isotropic stellar disks to local perturbations 
performed by Polyachenko et al. (\cite{polyachenko97b}) 
yields a critical value of $Q_{\rm s, crit}\approx 3.15$ for stellar disks 
with a flat rotation curve. Our numerical simulations indicate that the $m=2$ mode in 
the $Q_{\rm s}=3.15$ model~K5 saturates at a very low level of $C_2\approx 0.02-0.03$. 
As we shall see in \S~\ref{stabilization}, stellar disks characterized by such small 
values of $C_{\rm 2}$ can only develop a clumpy pattern with very small 
positive density perturbations rather than a spiral pattern or a bar. 


\subsection{Swing amplification} 
\label{swing} 
 
\begin{figure} 
   \resizebox{\hsize}{!}{ 
     \includegraphics[angle=0]{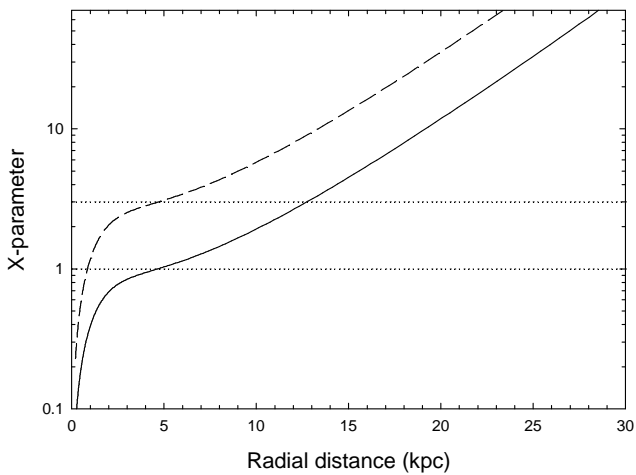}} 
   \caption {The initial radial distributions of the $X$-parameter for the 
   $m=1$ mode (dashed line) and the $m=2$ mode (solid line). Note that $X$ is identical 
   in all five models listed in Table~\ref{Table2}, because these models initially differ only in the values of 
   the velocity dispersion.} 
   \label{fig3} 
\end{figure}

An appealing physical interpretation for the growth of strong instabilities that are common 
in numerical simulations of differentially rotating stellar disks is swing amplification 
(Toomre \cite{toomre81}).  Amplification occurs when any leading disturbance unwinds 
into a trailing one due to differential rotation. 
It is helpful to introduce the parameter $X \equiv \lambda/\lambda_{\rm cr}$ 
when discussing the efficiency of swing amplification, where 
$\lambda \equiv 2\pi r/m $ is the circumferential wavelength of an $m$-armed disturbance 
and $\lambda_{\rm cr} \equiv 4\pi^2 G \Sigma/\kappa^2$ is the longest unstable wavelength 
in a cold disk.  According to Julian and Toomre 
(\cite{JT}), in a stellar disk with a flat rotation curve the gain of the swing amplifier 
is the largest when $1<X<3$. The gain also strongly depends on the value of the $Q$-parameter. 
Substantial amplification is also possible for $X<1$ if Q is safely below 2. 
 
\begin{figure} 
   \resizebox{\hsize}{!}{ 
     \includegraphics[angle=0]{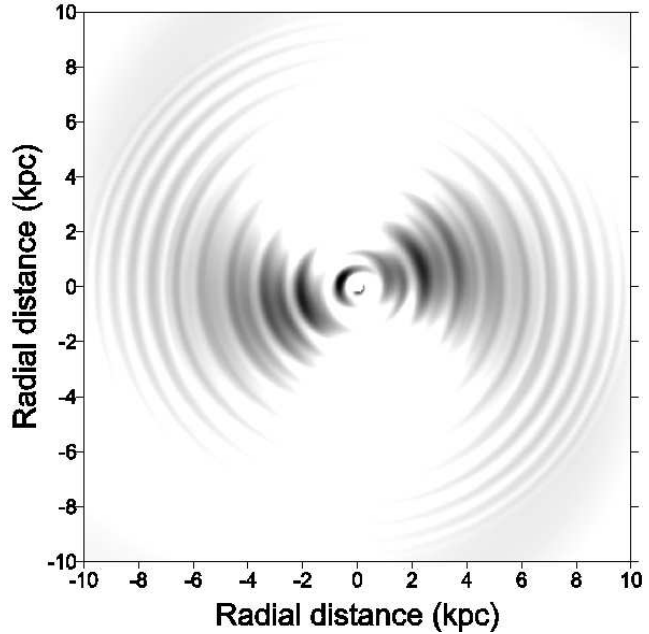}} 
   \caption {The positive density perturbation that develops in the inner 
   10~kpc region in model~K2 at $t=1.0$~Gyr. The lumpy structure is naturally 
   explained as the result of interference between leading and trailing short-wavelength disturbances 
   that propagate through the centre of the disk. The scale bar is in $M_\odot$~pc$^{-2}$.} 
   \label{fig4} 
\end{figure} 
 
Figure~\ref{fig3} shows the  initial 
radial distribution of the $X$-parameter for the $m=1$ (dashed line) and $m=2$ (solid line) 
disturbances. It is evident that the best conditions for swing amplification 
($1<X<3$) of the $m=1$ mode are met in the inner portion of the disk at 
($0.85-4.7$)~kpc, while the swing 
amplifier of the $m=2$ mode is maximal at intermediate radii ($4.7-12.8$)~kpc. 
However, feedback 
loops that turn leading disturbances into trailing ones 
must be present in stellar disks in order for swing amplification to destabilize 
the disk (see e.g. BT87 for a discussion). 
Since our stellar disks (initially) have no inner Lindblad resonances, 
the trailing disturbances can propagate through the centre and emerge on the other side 
as leading ones, thus providing a feedback for the swing amplifier. 
Figure~\ref{fig4} supports this point of view. 
It shows the positive density perturbation that develops in the inner 10~kpc in model~K2 
after 1~Gyr from the beginning of simulations. The lumpy structure that is seen in 
Fig.~\ref{fig4} is naturally explained as the result of interference between 
leading and trailing short-wavelength ($\lambda < \lambda_{\rm cr} $) disturbances, 
propagating through the disk centre. A similar phenomenon was seen in the numerical 
modelling by Toomre (\cite{toomre81}) and was discussed by BT87 (cf. Sect. 6.3, 2(c)). 
The $m=1$ density response dominates the innermost disk, while the $m=2$ density perturbation 
is the strongest in the intermediate and  outer regions in Fig.~\ref{fig4}. 
This tendency can indeed be expected from the radial distribution of the $X$-parameter 
for the $m=1$ and $m=2$ modes shown in Fig.~\ref{fig3}. 
The perturbation amplitudes gradually decline at larger radii. 
 
\begin{figure} 
   \resizebox{\hsize}{!}{ 
     \includegraphics[angle=0]{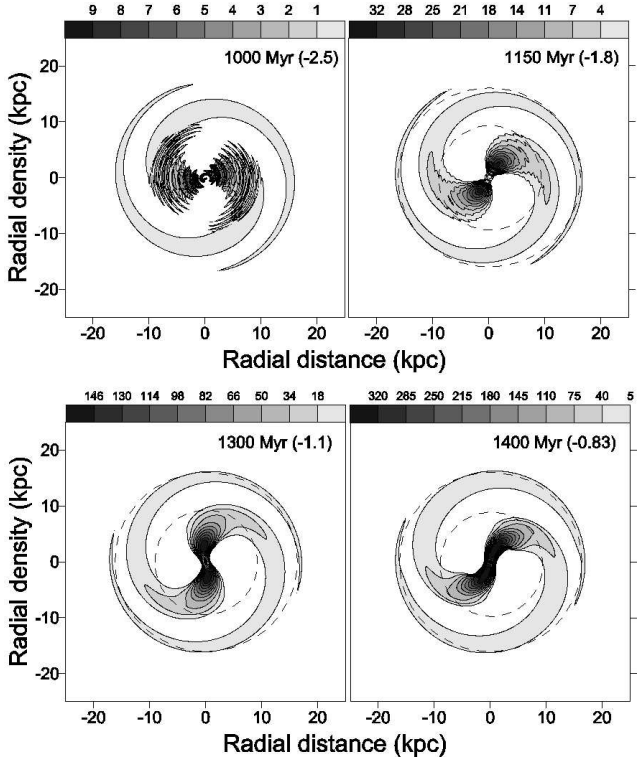}} 
   \caption {Positive density perturbations in model~K2 at four different 
   times indicated in each panel. The quantities in parentheses give the 
   values of $C_{\rm 2}(t)$ (in log units) at the corresponding evolutionary times. The 
   outer and inner dashed circles sketch the positions of corotation and 
   the outer Lindblad resonances. The scale bars give the positive density 
   perturbations in $M_\odot$~pc$^{-2}$.} 
   \label{fig5} 
\end{figure} 
 
To better illustrate the development of the $m=2$ instability, 
we plot in Fig.~\ref{fig5} the positive density perturbation obtained 
in model~K2 at four different times. 
The quantities in parentheses give the global Fourier 
amplitudes (in log units) at the corresponding times, while the scale bars give the 
amplitudes of positive density perturbations in $M_{\odot}$~pc$^{-2}$. 
It is clearly seen that the $m=2$ mode is the fastest growing mode. 
The lumpy structure at $t=1$~Gyr is 
gradually transforming into an $m=2$ spiral perturbation at $t=1.15$~Gyr. 
The spiral perturbation is mostly localized within the outer Lindblad resonance (OLR; shown by the outer 
dashed circle), which is in good agreement with the theoretical predictions.

\begin{figure} 
   \resizebox{\hsize}{!}{ 
     \includegraphics[angle=0]{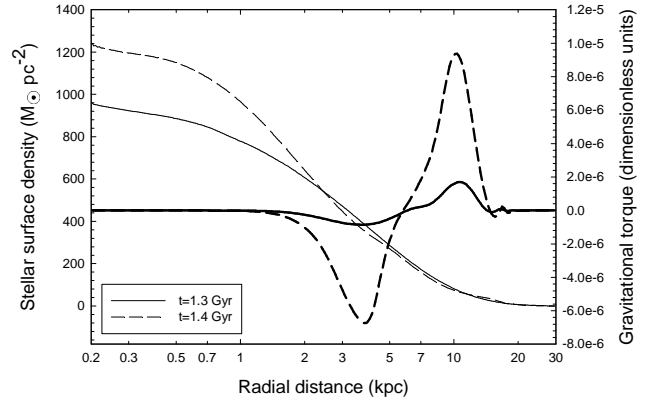} 
   } 
   \caption {The azimuthally averaged surface density distribution of stars 
   (thin solid and dashed lines) and the azimuthally averaged gravitational torque (thick 
   solid and dashed lines) obtained in model~K2 at two different evolutionary 
   times as indicated in the legend.} 
   \label{fig6} 
\end{figure}

\begin{figure} 
   \resizebox{\hsize}{!}{ 
     \includegraphics[angle=0]{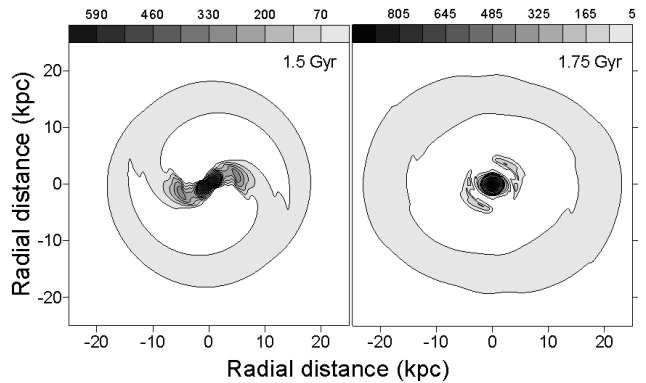} 
   } 
   \caption {Positive density perturbations in model~K2 at 1.5~Gyr (left 
   panel) and 1.75~Gyr (right panel).  The scale bar is in $M_\odot$~pc$^{-2}$.} 
   \label{fig7} 
\end{figure}

As the global Fourier amplitude $C_2(t)$ exceeds $0.1$ at $t\approx 1.3$~Gyr, the subsequent evolution of 
the $m=2$ spiral disturbance is governed by the non-linear effects of mass and angular 
momentum redistribution in the disk. Spiral disturbances are known to 
transfer mass inward and angular moment outward (see e.g. Laughlin et al. \cite{laughlin97}). 
The azimuthally averaged gravitational torque ${\Gamma}(r)$, which is 
the sum of the individual torques  $\tau=-m \, 
\partial \Phi /\partial \phi$ exerted on all computational cells at a given 
radius $r$, is known to be a good diagnostic tool for the 
mass and angular momentum redistribution in the disk. 
Figure~\ref{fig6} shows ${\Gamma}(r)$ and the azimuthally averaged surface density distribution 
in model~K2 obtained at $t=1.3$~Gyr and $t=1.4$~Gyr. 
The stars that are characterized by negative ${\Gamma}$ are losing 
angular momentum and spiralling into the centre, while the stars that are distinguished by 
positive ${\Gamma}$ are gaining angular momentum and moving radially outward. 
As a result, the density in the central region starts to grow considerably after 1.3~Gyr, 
which triggers the formation of a transient bar inside corotation. This process is evident in 
the right lower panel of Fig.~\ref{fig5}. 
The material that flows radially outward settles 
into a weak outer resonant ring near the position of the outer Lindblad resonance at approximately 
15~kpc (note that ${\Gamma}$ becomes zero outside the outer Lindblad resonance).

A shortage  of matter that develops in the intermediate region weakens 
the spiral structure and terminates the mass and angular momentum redistribution. 
In the end, the bar transforms into a compact dense central disk, 
while the spiral structure virtually disappears. The outer resonant ring 
occupies a substantial portion of the outer disk. 
Figure~\ref{fig7} illustrates this process and shows the positive density 
perturbation that develops in model~K2 at $t=1.5$~Gyr (left) and $t=1.75$~Gyr 
(right). This phenomenon is a nice example of 
the dominant mode self-interaction originally studied by Laughlin et al. (\cite{laughlin97}) 
in the context of gaseous disks.

\subsection{Stabilization at larger $Q_{\rm s}$} 
\label{stabilization} 
 
\begin{table} 
\caption{Pattern speeds and position of resonances} 
\label{Table3} 
\vskip 0.1cm 
\begin{tabular}{lllc} 
\hline 
Model & $\Omega_{\rm p}^{(a)}$ & CR & OLR \\ [2 pt] 
\hline 
K1 & 30.2 & 6.2 & 11.5 \\ 
K2 & 23.1 & 8.4 & 15 \\ 
K3 & 19.1 & 10.4 & 18.1 \\ 
K4 & 11.5 & 17.6 & 24.1 \\ 
\hline 
\end{tabular} 
 
\medskip 
$^{(a)}$ the pattern speeds  are in km~s$^{-1}$~kpc$^{-1}$ and positions 
of corotations (CR) and outer Lindblad resonances (OLR) are in kpc. 
 
\end{table} 
 
The stabilizing effect of increasing $Q_{\rm s}$ is clearly seen in Fig.~\ref{fig2}. 
The $m=2$ perturbations in hotter disks are characterized by lower growth rates 
and they take longer time to saturate. 
The global Fourier amplitudes in the saturation phase 
are noticeably larger in models K1--K3 ($C_2(t) \approx 0.1-0.25$) than in models K4 and K5 
($C_2(t) \approx 0.02-0.05$), indicating a weaker instability for hotter disks. 
Figure~\ref{fig8} 
shows the positive density perturbation that develops in model~K4 in the 
saturation phase at 6.8~Gyr (left panel) and 7.0~Gyr (right panel). The bar and 
the spiral arms have a patchy, flocculent appearance. The positive density 
perturbation  (in physical units of $M_\odot$ pc$^{-2}$) in the saturation 
phase is on average {\it an order of magnitude smaller} than in model~K2 (Fig.~\ref{fig5}). 
Clearly, the growth of the global modes is considerably suppressed in model~K4. 
Figure~\ref{fig9} 
shows the positive density perturbation that develops in model~K5 in the 
saturation phase at 7.6~Gyr (left panel) and 8.8~Gyr (right panel). 
It is obvious that the positive density perturbation in model~K5 is characterized by a 
clumpy structure of low amplitude rather than by a regular spiral or bar. 
This is because in model~K5  both the $m=1$ and $m=2$ modes 
saturate at a nearly equal (and very small) value of $C_{\rm 1,2}\approx 0.02-0.03$. 
Thus, we confirm the theoretical predictions of Polyachenko et al. (\cite{polyachenko97b}), 
who have analytically shown that stellar disks characterized by $Q_{\rm s}< 3.15$ and a flat rotation curve are 
unstable to local density perturbations.

\begin{figure} 
   \resizebox{\hsize}{!}{ 
     \includegraphics[angle=0]{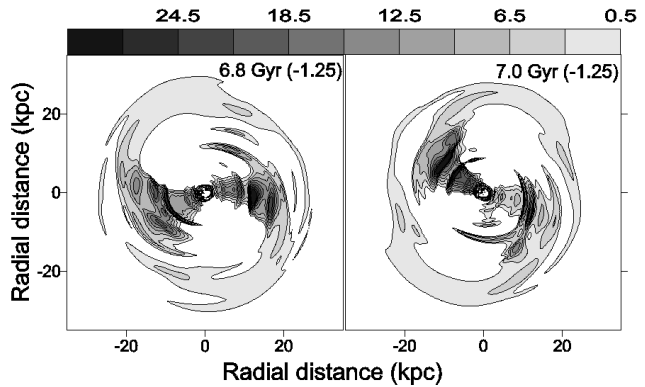} 
   } 
   \caption {Positive density perturbations in model~K4 at 6.8~Gyr (left 
   panel) and 7.0~Gyr (right panel).   
   The quantities in parentheses give the 
   values of $C_{\rm 2}(t)$ (in log units) at the corresponding evolutionary times. 
   Note a patchy, flocculent structure 
   of the bar and spiral arms. The scale bar is in $M_\odot$~pc$^{-2}$.} 
   \label{fig8} 
\end{figure} 
 
\begin{figure} 
   \resizebox{\hsize}{!}{ 
     \includegraphics[angle=0]{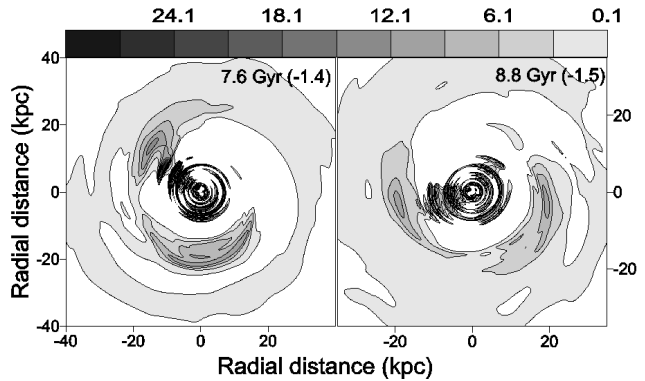} 
   } 
   \caption {The same as in Fig.~\ref{fig8} but for model~K5.} 
   \label{fig9} 
\end{figure} 
 
 
We have shown in Sect.~\ref{swing} 
that the physical interpretation for the instability growth in our models 
is swing amplification, with the feedback provided by the trailing disturbances 
propagating through the disk centre. Julian and Toomre (\cite{JT}) have shown that 
the gain of swing amplification is greatly sensitive to the values of the $X$- and 
$Q$-parameters. More specifically, the gain diminishes considerably for $Q_{\rm s}>2$. 
Since the initial radial profiles of the $X$-parameter are identical in all model disks 
(see Fig.~\ref{fig3}), the increase in $Q_{\rm s}$ and, by implication, in 
random motions of stars suppresses the short-wavelength disturbances and thus inhibits the 
swing amplifier. 
 
The comparison of dynamical properties of our model disks suggests 
another stabilization effect that may be present in hotter stellar disks. 
Table~\ref{Table3} gives the pattern speeds and the positions of corotation 
and outer Lindblad resonances in models K1--K4. 
These values are computed in the saturation phase and may slightly change during the evolution. 
It is clearly seen that models with larger $Q_{\rm s}$ are distinguished by a slower pattern speed 
$\Omega_{\rm p}$. 
This is in qualitative agreement with the predictions of the linear stability analysis 
of round galactic disks by Pichon \& Cannon (\cite{Pichon}), who have shown that 
colder disks yield more centrally concentrated and faster rotating spirals. 
Slower pattern speeds in hotter disks imply that the latter can easier 
develop the inner Lindblad resonance due to the rearrangement of mass and angular momentum. 
Indeed, solid lines in Fig.~\ref{fig10} show the initial radial 
profiles of stellar angular velocity $\Omega$ and quantities $\Omega \pm 
\kappa/2$, while the dashed lines give the pattern speeds $\Omega_{\rm 
p}$ in models K1--K4 (from top to bottom). None of the models 
have the inner Lindblad resonance but the tendency is clear: hotter disks have better 
chances for the inner Lindblad resonance to occur. The inner Lindblad resonance would prohibit 
the propagation of spiral disturbances through the disk centre, 
thus providing the cutoff for the swing amplification feedback.

\begin{figure} 
   \resizebox{\hsize}{!}{ 
     \includegraphics[angle=0]{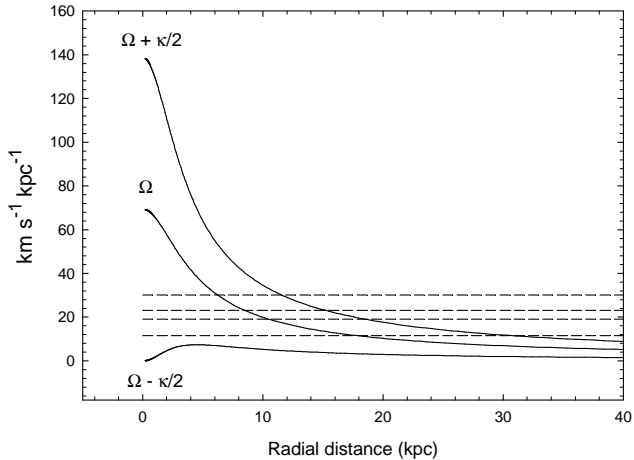} 
   } 
   \caption {Radial profiles of the stellar angular velocity $\Omega$ (dotted-dashed 
      line) and quantities $\Omega \pm k/2$. The pattern speeds $\Omega_{\rm 
      p}$ of models K1--K4 are shown by the dashed lines (from top to bottom).} 
   \label{fig10} 
\end{figure}

\subsection{The $m=1$ density perturbation} 
 
The initial radial distributions of the $X$-parameter in Fig.~\ref{fig3} (identical for 
all our model disks) suggest that the $m=1$ perturbations should dominate in the innermost 
parts of stellar disks. In fact, this can already  be seen in Fig.~\ref{fig4}, where the 
positive density response in the inner 1~kpc has a crescent shape. 
We find that hotter disks are more strongly disposed to the development 
of the $m=1$ density perturbations than colder disks. 
Although the instability in disks with $Q_{\rm 
s} \la 2.5$ is dominated by the symmetric $m=2$ mode, the 
lopsided $m=1$ mode in disks with $Q_{\rm s}> 2.5$ may compete 
with or even prevail over the $m=2$ mode. 
This tendency is clearly seen in Fig.~\ref{fig11}, in which we show the temporal evolution of the 
$m=1$ mode (dashed lines) and $m=2$ mode (solid lines) in models~K3--K6. Model~K6 is 
initially characterized by $Q_{\rm s}=3.5$. It is obvious that the $m=2$ mode dominates during the evolution 
(except for the very early phase) in models with $Q_{\rm s}=1.6$ (K3) and $Q_{\rm s}=2.5$ (K4). 
This tendency is not seen in the $Q_{\rm s}=3.15$ model~K5, in which both the $m=1$ and $m=2$ 
modes have a nearly equal 
global Fourier amplitudes and growth rates in the saturation phase. 
Finally, the $Q_{\rm s}=3.5$ 
model~K6 is characterized by the dominant $m=1$ mode. 
 
The preponderance of the $m=1$ mode in model~K6 is due to 
the fact that one-armed perturbations are much more difficult to stabilize 
(Evans \& Read \cite{Evans}). Indeed, the $m=1$ modes have no inner Lindblad 
resonance (the quantity $\Omega-\kappa$ is negative), which removes a powerful stabilizing 
effect for the $m=1$ mode. On the other hand, the stabilizing effect of the inner Lindblad resonance 
for the $m=2$ mode is expected to become stronger along the sequence of increasing $Q_{\rm s}$ 
(see Fig.~\ref{fig10}). As a consequence, the strength of the $m=2$ mode in the saturation phase 
(as determined by $C_2(t)$) 
decreases for hotter disks, whereas the strength of the $m=1$ mode (as determined by $C_1(t))$ 
stays nearly constant in all four models shown in Fig.~\ref{fig11}. 
We note that according 
to Evans \& Read (\cite{Evans}) stellar disks with a sharp central cut-out (stellar density drops 
to zero in the centre) are much more unstable to the development of the $m=1$ modes. 
 

\begin{figure} 
   \resizebox{\hsize}{!}{ 
     \includegraphics[angle=0]{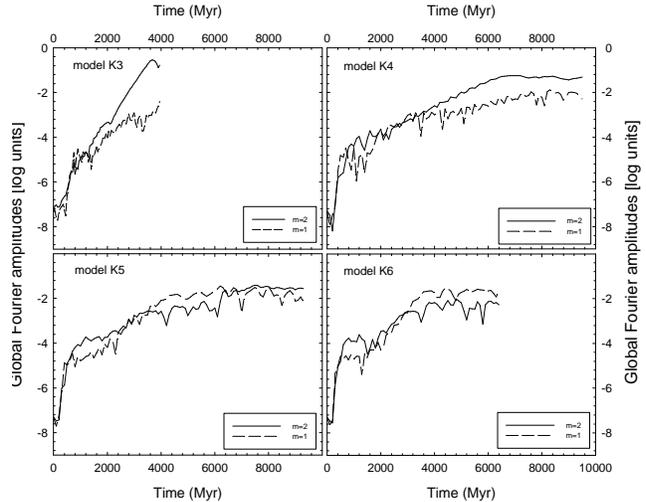}} 
   \caption {Temporal evolution of the $m=1$ modes (dashed lines) and the $m=2$ modes 
   (solid lines) in models K3, K4, K5, and K6. Model K6 is characterized by $Q_{\rm s}=3.5$. 
   Note that in the saturation phase the global Fourier amplitudes of the $m=2$ modes  decrease along the sequence of increasing 
   $Q_{\rm s}$, while the global Fourier amplitudes of the $m=1$ modes stay nearly constant in all 
   four models. 
   } 
   \label{fig11} 
\end{figure} 
 

 
\subsection{Vertex deviation} 
 
The non-isotropic nature of the BEADS-2D code allows for a numerical study of the 
velocity ellipsoid properties in spiral galaxies. 
The non-axisymmetric component of the spiral gravitational field may produce 
considerable perturbations on stellar orbits. 
A vertex deviation $l_{\rm v}$ 
can be used to study the magnitude of the spiral gravitational field 
(Kuijken \& Tremaine \cite{KT}). 
The vertex deviation has a simple geometric interpretation: it is the angle between 
the major axis of the velocity ellipsoid (the surface that has semi-axes 
$\sigma_{zz}$, $\sigma_{rr}$, and $\sigma_{\phi\phi}$) 
at a given position in the disk and 
the centre-anticentre direction. Based on the above explanation, the value 
of $l_{\rm v}$ can be defined as (Binney and Merrifield \cite{binney98}, 
p.\ 630) 
\begin{equation} 
l_{\rm v}={1 \over 2} {\rm atan} \left( {2 \sigma^2_{r\phi} \over \sigma^2_{rr}- 
\sigma^2_{\phi\phi}}   \right). 
\label{vertex1} 
\end{equation} 
This definition is incomplete because it implicitly assumes 
the epicycle approximation, i.e $\sigma^2_{rr}>\sigma^2_{\phi\phi}$. 
As a result, the value of $l_{\rm v}$ is limited to the $[-45^\circ , 45^\circ]$ 
range depending on the sign of $\sigma^2_{r\phi}$. 
As a matter of fact, the epicycle approximation may break down in the presence 
of a strong non-axisymmetric gravitational field of spiral arms. 
Consequently, the azimuthal dispersion may become (locally) larger than the radial 
one and the angle between the major axis of the velocity ellipsoid 
and the centre-anticentre direction may exceed $\pm 45^\circ$. To take 
this possibility into account, we extend the definition of 
the classical vertex deviation as follows: 
 
\[\tilde{ l}_{\rm v} = \left\{ \begin{array}{ll} l_{\rm v} 
    & \,\, \mbox{if $\sigma^2_{rr} > \sigma^2_{\phi\phi}$} \\ 
     l_{\rm v} + {\rm sign}\left(\sigma^2_{r\phi}\right) \cdot \displaystyle\frac{\pi}{2} 
    & \,\, \mbox{if $\sigma^2_{rr}<\sigma^2_{\phi\phi}$}.   \end{array} 
   \right. \] 
 
Black dashes in Fig.~\ref{fig12} show the major axes of the velocity ellipsoids 
superimposed on the positive density perturbation map which is obtained in model~K2 
at $t=1.4$~Gyr. 
Considerable vertex deviations  are seen near the convex edges of spiral perturbations 
and near the bar, where $\tilde{l}_{\rm v}$ may become as high as $89^\circ$. 
Interestingly, the concave edges of spiral perturbations do not produce noticeable 
vertex deviations. The outer axisymmetric regions and the inter-arm regions   
are also characterized by near-zero vertex deviations. 
The temporal evolution of a mass-weighted value of $\tilde{l}_{\rm v}$ over 
the entire disk is approximately correlated with $C_2(t)$. 
For instance $l_{\rm v}=1.8^\circ$ at $t=1.0$~Gyr and $l_{\rm v}=12.6^\circ$ at $t=1.4$~Gyr. 
 
Since we start our simulations with a zero vertex deviation ($\sigma_{r\phi}=0$), we confirm that 
the spiral gravitational field produces a vertex deviation -- a conclusion 
previously made by Kuijken \& Tremaine (\cite{KT}) on analytical grounds. 
A more detailed study of the properties of the velocity ellipsoids in spiral galaxies 
will be presented in a forthcoming paper. 
 
In contrast to gaseous disks, stellar disks are essentially non-isotropic. 
In the epicycle approximation, the ratio of radial to azimuthal components of the velocity ellipsoid 
can be described by equation (\ref{epic}), implying that stellar disks are 
colder in the azimuthal direction than in the radial direction. 
The initial ratios  $\sigma_{\phi\phi}:\sigma_{rr}$ in model~K2 lie in the ($1-0.71$) range, 
with the minimum and maximum values being 
near the centre and at the outer edge of the stellar disk, respectively. 
As the disk evolves and the spiral structure emerges, the ratios $\sigma_{\phi\phi}:\sigma_{rr}$ 
start to show considerable deviations from the initial values. 
In the saturation phase, the bulk of the stellar disk is characterized by $\sigma_{\phi\phi}:\sigma_{rr}$ 
lying in the $0.65-0.75$ range, 
which is comparable to a measured mean value of $\sigma_{\phi\phi}:\sigma_{rr}\approx 0.8$ 
for the disk of NGC~488 (Gerssen et al.\ \cite{Gerssen}). 
However, considerably smaller values of $\sigma_{\phi\phi}:\sigma_{rr} \la 
0.4$ are found at the leading edges of spiral arms, implying that the epicycle approximation 
breaks down there. A more detailed study of the anisotropy of 
the velocity ellipsoid will be presented in a future paper. 
 
 
\begin{figure} 
   \resizebox{\hsize}{!}{ 
 \includegraphics{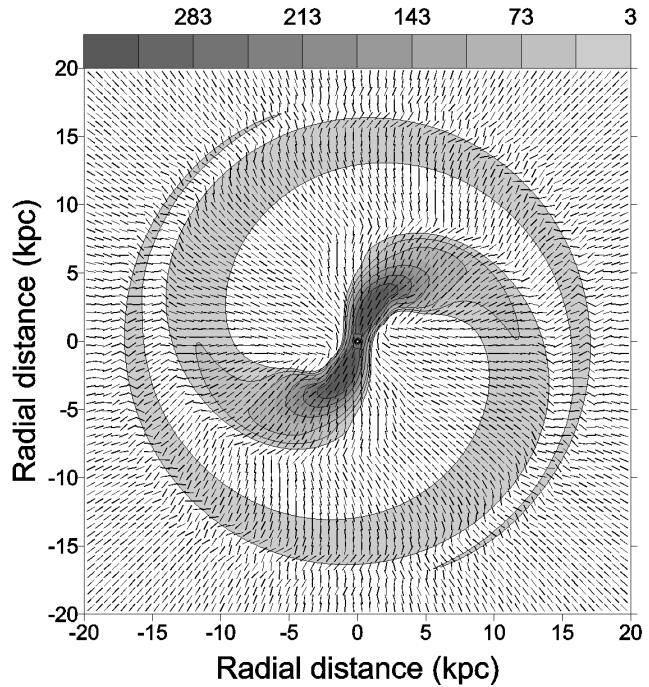}} 
 \caption{The major axes of velocity ellipsoids superimposed 
      on the surface density distribution of the kinetic model K2 ($Q_s = 1.3$) at the 
      time $t=1.3~\mathrm{ Gyr}$. The angle between the major axis of 
      velocity ellipsoid and the centre-anticentre direction ($\phi=const$) 
      gives a value of vertex deviation.} 
 \label{fig12} 
\end{figure} 
 
%
%
 
\subsection{Miscellaneous} 
In all models considered in this paper, the outer reflecting boundary is 
set at a large enough distance ($r\ga 30$~kpc) so as to exclude its 
influence on spiral generation and growth. A factor of 2 variation 
in the position of the inner reflecting boundary ($0.2$~kpc in the present simulation) 
does not introduce a noticeable change in the numerical results. 
The radial and azimuthal components of velocity dispersions may accidentally 
become negative in the late saturation phase, probably due to large gradients 
of the stellar potential. In that case the time step is reduced 
and the solution is sought again. In the rare case that this procedure 
fails to resolve the problem, we set the velocity dispersion in particular 
computational cells to a predefined small value.

Our simulations have shown that artificial viscosity is usually not needed 
in models with $Q_{\rm s} \ga 1.3$. 
However, the artificial viscosity has to be included 
if the evolution of spiral instabilities is followed deeply into the nonlinear regime 
in models with $Q_{\rm s}<1.3$. 
The solution is only slightly modified by the artificial viscosity. For 
instance, the relative difference in the growth rate of the dominant $m=2$ mode in model~K1 
with and without artificial viscosity is kept below $2\%$ during the simulations. 
 

 
\section{Summary and Conclusions} 
\label{sect_summary}

We have developed the Boltzmann moment equation approach for the 
dynamics of stars (BEADS-2D), which is a finite-difference Eulerian 
code based on the Boltzmann moment equations up to second 
order. We adopt the zero-heat-flux approximation to close the system 
of moment equations. The BEADS-2D code is formulated in the polar 
coordinates $(r,\phi)$, which makes it most suitable for the numerical modelling 
of flattened galactic stellar disks. By calculating the tensor of the 
velocity dispersions, we can follow the anisotropy and the vertex deviation of the 
stellar system. Compared to numerical codes assuming a polytropic 
equation of state (see e. g. Korchagin et al. \cite{korchagin00}),   
the BEADS-2D code allows for a larger variety of simulations, because the 
mass and velocity dispersion profiles can be chosen independently. 
For the reader's convenience, we provide the full set of 
three-dimensional Boltzmann moment equations up to second order in Appendix~A . These 
equations are formulated in the cylindrical coordinates $(z,r,\phi)$ 
using the usual zero-heat-flux approximation and are suitable for a 
global modelling of galactic disks without simplifying geometry 
assumptions. 
 
As an example of the utility of the BEADS-2D code, we study the time-dependent 
evolution of exponential stellar disks 
characterized initially by different values of the Toomre parameter $Q_{\rm s}$. 
The disks are embedded in a static dark matter halo, yielding a   
nearly flat rotation curve (except for the innermost several kiloparsecs 
characterized by rigid rotation). 
Specifically, we find the following. 
 
 
(i) We confirm the results of linear stability analysis by 
Polyachenko et al. (\cite{polyachenko97b}), who obtained a stability threshold 
value of $Q_{\rm s} \approx 3.15$ for stellar disks with a {\it purely} 
flat rotation curve. 
A physical interpretation for the instability is 
swing amplification of short-wavelength trailing disturbances, which 
propagate through the disk centre and provide a feedback mechanism for 
the swing amplifier. 
 
(ii) The character of the instability is distinct in the linear and 
nonlinear regimes, the tentative boundary between which we define by 
the global Fourier amplitude $C_2(t)=0.1$. 
As a reference model, we investigated a stellar disk with $Q_{\rm s}=1.3$. 
In the early linear phase, 
instability demonstrates a typical lumpy pattern, which grows in amplitude 
with time and gradually 
transformes in to a global two-armed spiral in the late linear phase. 
This lumpy pattern results most probably from the interference between leading and 
trailing disturbances, propagating through the disk centre. A similar 
phenomenon is discussed in Binney \& Tremaine (\cite{binney87}}) 
in connection with the numerical simulations of Toomre (\cite{toomre81}). 
In the nonlinear phase, the redistribution of 
mass and angular momentum by the gravitational torques of spiral arms 
comes into play. The gravitational torques produce an inflow of mass, which 
forms  a transient central bar inside corotation, 
and an outflow of mass, which forms a ring at the OLR. 
This process of mass redistribution continues until the two-armed spiral is weakened 
by the shortage of matter. Finally, a transient bar is transformed into 
the compact dense central disk and the spiral arms virtually disappear, 
leaving only the outer diffuse ring at the OLR.

(iii) The instability in disks with $Q_{\rm s} \la 2.5$ is mostly 
dominated by the symmetric $m=2$ mode, while the 
lopsided $m=1$ mode in disks  with $Q_{\rm s}> 2.5$ may compete 
with or even prevail over the $m=2$ mode. This is due the fact that the $m=1$ mode 
has no inner Lindblad resonance, which makes it difficult to stabilize. 
On the other hand, the stabilizing effect of the inner Lindblad resonance on the 
$m=2$ mode is expected to grow along the sequence of increasing $Q_{\rm s}$. 
 
 
(iv) Stellar disks near a stability limit of $Q_{\rm s}=3.15$ 
still  develop a clumpy structure.  However, 
both the growth timescales (comparable to the Hubble time) 
and the small amplitudes of positive density perturbations 
make such instabilities difficult to observe at best. 
 
(v) Stellar disks are stabilized by one or more of the 
following procedures. a) An 
increase in the stellar velocity dispersions suppresses the growth of 
short-wavelength density perturbations and thus inhibits the swing 
amplifier.  b) Slower pattern speeds in hotter disk make it more 
likely for the inner Lindblad resonance to occur, thus providing the 
cutoff for the sing amplification feedback.

The presence of the non-diagonal velocity dispersion terms in the 
Boltzmann moment equations allows us to study the vertex deviation. In 
most regions of the stellar disk the velocity ellipsoid is well 
aligned with the radial axes of the polar grid. 
In agreement 
with analytical predictions by Kuijken \& Tremaine (\cite{KT}), 
considerable vertex deviations show up in regions with strongly 
perturbed mass distributions, i.e.\ near the spiral arms.  The vertex 
deviations are large at the convex edge of the spiral 
arms, whereas they are small at the concave edge.  The mean vertex 
deviations correlate well with the global Fourier amplitudes. 
Near the convex edge of the spiral arms, the ratio of radial to azimuthal components of the velocity ellipsoid 
can deviate considerably from 
the values predicted from the epicycle approximation. 
A more detailed numerical study of the properties of velocity ellipsoids 
in spiral galaxies will be presented in a future paper.


\section*{Acknowledgments} 
   We would like to thank the referee for several important suggestions, 
   especially for pointing out the motivation of this paper more clearly. 
   The authors are grateful to Prof.\ Shantanu Basu for carefully reading 
   the manuscript, correcting the English language usage, and suggesting stylistical   
   improvements. 
   A research visit of EIV has been supported by a grant 
   of the University of Vienna. EIV also gratefully acknowledges present support 
   from a CITA National Fellowship. Our special thanks are to Prof. Martin Houde 
   for generously providing computational facilities.


 
\appendix 
 
\section{The 3D-Boltzmann moment equations up to second order} 
 
  The Boltzmann moment equations (up to second order) read in 
cylindrical coordinates $(r,\phi,z)$ as: \\ 
\noindent 
{\bf Continuity equation:} 
\begin{equation} 
     {\partial \rho \over \partial t} 
  + {1 \over r} {\partial \over \partial r} (r\cdot \rho \cdot u_{r}) 
  + {1 \over r} {\partial \over \partial \phi} (\rho \cdot u_{\phi}) 
  + {\partial \over \partial z} (\rho \cdot u_{z}) = 0. 
\label{eq_3dkin_rho} 
\end{equation} 
 
\noindent 
{\bf Momentum equations:} 
\begin{eqnarray} 
    {\partial \over \partial t}(\rho u_{z}) 
 &+& {1 \over r} {\partial \over \partial r}(r \cdot\rho u_{z} \cdot u_{r}) 
  + {\partial \over \partial z} (\rho u_{z}\cdot u_{z})  \nonumber \\ 
 &+& {1 \over r} {\partial \over \partial \phi} (\rho u_{z}\cdot u_{\phi}) 
  + {\partial \over \partial z} (\rho \sigma_{zz}^2) \nonumber \\ 
 &+&{1 \over r} {\partial \over \partial r} (r\cdot \rho\sigma_{zr}^2) 
  +{1 \over r}{\partial \over \partial \phi} (\rho \sigma_{z \phi}^2) 
  + \rho {\partial \Phi \over \partial z}=0, 
\end{eqnarray} 
 
\begin{eqnarray} 
   {\partial \over \partial t}(\rho u_{r}) 
 &+& {\partial \over \partial z} (\rho u_{r} \cdot u_{z}) 
  +{1 \over r} {\partial \over \partial r}(r\cdot\rho u_{r} \cdot u_{r}) \nonumber \\ 
 &+& {1 \over r}{\partial \over \partial \phi} (\rho u_{r}\cdot  u_{\phi}) 
  - {\rho u_{\phi}^2 \over r}  + {\partial \over \partial z}(\rho \sigma_{rz}^2) \nonumber \\ 
 &+& {1\over r}{\partial \over \partial r}(r \cdot \rho\sigma_{rr}^2) 
  + {1 \over r}{\partial \over \partial \phi}(\rho \sigma_{r \phi}^2) 
  - {\rho \sigma_{\phi\phi}^2 \over r} \nonumber\\ 
 &+& \rho {\partial \Phi \over \partial r} =0, 
\end{eqnarray} 
 
\begin{eqnarray} 
    {1 \over r} \left[ {\partial \over \partial t} (\rho r u_{\phi })\right. 
  &+& \left. {1\over r} {\partial \over \partial r} (r \cdot \rho r u_{\phi}\cdot u_{r }) 
   +  {1\over r } {\partial \over \partial \phi} (\rho r u_{\phi}\cdot u_{\phi}) \right. \nonumber \\ 
  &+& \left. {\partial \over \partial z} (\rho r u_{\phi}\cdot u_{z}) \right] 
   + {\partial \over \partial z} (\rho \sigma_{\phi z}^2) \nonumber\\ 
  &+& {\partial \over \partial r} (\rho \sigma_{r \phi}^2) 
   + {1\over r}{\partial \over \partial \phi} (\rho \sigma_{\phi \phi}^2) 
   + {2\over r} \rho \sigma_{r \phi}^2 \nonumber \\ 
  &+&\rho {1\over r}{\partial \Phi \over \partial \phi}=0. 
\end{eqnarray}

\noindent 
{\bf Velocity dispersion equations:} 
\begin{eqnarray} 
      {\partial \over \partial t} (\rho \sigma_{zz}^2) 
  &+& {1\over r}{\partial \over \partial r} (r\cdot \rho \sigma_{zz}^2 \cdot u_{r}) 
   + {\partial \over \partial z} (\rho \sigma_{zz}^2 \cdot u_{z}) \nonumber \\ 
  &+&  {1\over r}{\partial \over \partial \phi} (\rho \sigma_{zz}^2\cdot u_{\phi}) 
   + 2 \rho \sigma_{rz}^2 {\partial u_{z} \over \partial r} 
   + 2 \rho\sigma_{zz}^2 {\partial u_{z} \over \partial z} \nonumber \\ 
  &+& 2 \rho \sigma_{z\phi}^2 {1\over r} {\partial u_{z} \over \partial \phi} =0, 
\end{eqnarray}

\begin{eqnarray} 
      {\partial \over \partial t} (\rho \sigma_{rr}^2) 
  &+& {1\over r}{\partial \over \partial r}(r\cdot \rho \sigma_{rr}^2\cdot u_{r}) 
   + {\partial \over \partial z} (\rho \sigma_{rr}^2\cdot u_{z}) \nonumber \\ 
  &+&{1\over r} {\partial \over \partial\phi}(\rho \sigma_{rr}^2\cdot u_{\phi}) 
   + 2\rho \sigma_{rr}^2 {\partial u_{r}\over \partial r} 
   +  2\rho \sigma_{rz}^2 {\partial u_{r} \over \partial z} \nonumber\\ 
  &+& 2 \rho \sigma_{r \phi}^2 {1\over r}{\partial u_{r} \over \partial\phi} 
   - {4 \rho \sigma_{r\phi}^2  u_{\phi}\over r }=0, 
\end{eqnarray} 
 
\begin{eqnarray} 
      {\partial \over \partial t} (\rho \sigma_{\phi\phi}^2) 
  &+& {1\over r}{\partial \over \partial r} (r\cdot\rho \sigma_{\phi\phi}^2\cdot 
       u_{r}) 
   +{\partial \over \partial z} (\rho \sigma_{\phi \phi}^2\cdot u_{z})\nonumber \\ 
  &+& {1 \over r}{\partial \over \partial \phi} (\rho \sigma_{\phi\phi}^2 \cdot u_{\phi}) 
   + {{2 \rho \sigma_{\phi \phi}^2} u_{r} \over r} + {2 \rho u_{\phi} \sigma_{r\phi}^2 \over r} \nonumber\\ 
  &+& 2 \rho \sigma_{r \phi}^2 {\partial u_{\phi} \over \partial r} 
   +  2 \rho \sigma_{z \phi}^2 {\partial u_{\phi} \over \partial z} \nonumber \\ 
  &+& 2 \rho \sigma_{\phi\phi}^2 {1\over r}{\partial u_{\phi} \over \partial \phi}  =0. 
\end{eqnarray}

\noindent 
{\bf Off-diagonal elements of the velocity-dispersion tensor:} 
\begin{eqnarray} 
      {\partial \over \partial t} (\rho \sigma_{r \phi}^2) 
  &+& {1\over r}{\partial \over \partial r} (r\cdot \rho \sigma_{r\phi}^2\cdot 
          u_{r}) 
   + {\partial \over \partial z} (\rho \sigma_{r \phi}^2\cdot u_{z})\nonumber\\ 
  &+& {1 \over r} {\partial \over \partial \phi} (\rho \sigma_{r\phi}^2 \cdot 
          u_{\phi}) 
   + {\rho \sigma_{r \phi}^2 u_{r} \over r } 
   + {\rho \sigma_{rr}^2  u_{\phi} \over r} \nonumber \\ 
  &-& {2 \rho \sigma_{\phi \phi}^2 u_{\phi}\over r} 
   + \rho \sigma_{rr}^2 {\partial u_{\phi} \over \partial r} 
   + \rho \sigma_{r \phi}^2 {\partial u_{r} \over \partial r} \nonumber \\ 
  &+& \rho \sigma_{r z}^2 {\partial u_{\phi} \over \partial z}   
   + \rho \sigma_{z \phi}^2 {\partial u_{r} \over \partial z} 
   + \rho \sigma_{\phi \phi}^2 {1\over r}{\partial u_{r} \over \partial \phi}\nonumber \\ 
  &+&   \rho \sigma_{r\phi}^2 {1\over r} {\partial u_{\phi} \over \partial \phi}=0, 
\end{eqnarray}

\begin{eqnarray} 
      {\partial \over \partial t} (\rho \sigma_{z\phi}^2) 
   &+& {1\over r}{\partial \over \partial r} (r\cdot\rho \sigma_{z\phi}^2\cdot 
          u_{r}) 
    + {\partial \over \partial z} (\rho \sigma_{z \phi}^2 \cdot u_{z}) \nonumber\\ 
   &+& {1 \over r} {\partial \over \partial \phi} (\rho \sigma_{z\phi}^2 \cdot 
          u_{\phi}) 
    + {\rho \sigma_{z \phi}^2 u_{r} \over r} 
    + {\rho \sigma_{rz}^2 u_{\phi} \over r} \nonumber \\ 
   &+& \rho \sigma_{rz}^2 {\partial u_{\phi} \over \partial r} 
    + \rho \sigma_{r\phi}^2 {\partial u_{z} \over \partial r} 
    + \rho \sigma_{zz}^2 {\partial u_{\phi} \over \partial z} \nonumber \\ 
   &+& \rho \sigma_{z \phi}^2 {\partial u_{z} \over \partial z} 
    + \rho \sigma_{\phi \phi}^2 {1\over r} {\partial u_{z} \over \partial \phi} \nonumber \\ 
   &+& \rho \sigma_{z \phi}^2 {1 \over r} {\partial u_{\phi} \over \partial \phi} =0, 
\end{eqnarray} 
 
\begin{eqnarray} 
      {\partial \over \partial t} (\rho \sigma_{rz}^2) 
   &+& {1\over r}{\partial \over \partial r} (r\cdot \rho \sigma_{rz}^2\cdot 
        u_{r}) 
    +  {\partial \over \partial z} (\rho \sigma_{rz}^2\cdot u_{z})\nonumber\\ 
   &+& {1\over r}{\partial \over \partial \phi}( \rho \sigma_{zr}^2 
        \cdot u_{\phi}) 
    - {2 \rho \sigma_{z \phi}^2 u_{\phi} \over r} 
    + \rho \sigma_{rr}^2 {\partial u_{z} \over \partial r} \nonumber\\ 
   &+& \rho \sigma_{zz}^2 {\partial u_{r} \over \partial z} 
    + \rho \sigma_{rz}^2 {\partial u_{z} \over \partial z} 
    + \rho \sigma_{r\phi}^2 {1\over r}{\partial u_{z} \over \partial \phi}\nonumber \\ 
   &+& \rho \sigma_{z\phi}^2 {1\over r} {\partial u_{r} \over \partial \phi} 
    + \rho \sigma_{zr}^2 {\partial u_{r} \over \partial r} =0. 
    \label{eq3d_sigma_rz} 
\end{eqnarray} 
Note that terms of third or higher order are neglected (zero-heat-flux approximation). 
 
\section{The six components of the artificial 
viscosity stress tensor $\bl Q$} 
 
  Here we give the components of the artificial viscosity stress 
tensor used in our simulations.

\begin{eqnarray} 
Q_{zz}&=&2 \mu_{\rm v} \left( {\partial u_{z} \over \partial z}  - {1\over 3} 
(\bl \nabla \cdot \bl u) \right) \\ 
Q_{zr}&=& \mu_{\rm v} \left( {\partial u_{z} \over \partial r} + 
{\partial u_{r} \over \partial z} \right) \\ 
Q_{z\phi}&=& \mu_{\rm v} \left( {\partial u_{\phi} \over \partial z} + 
{1\over r} {\partial u_{z} \over \partial \phi}  \right) \\ 
Q_{rr}&=& 2\mu_{\rm v} \left( {\partial u_{r} \over \partial r} - 
{1\over 3} (\bl \nabla \cdot \bl u) \right) \\ 
Q_{r\phi} &=& \mu_{\rm v} \left( {1\over r} {\partial u_{r} \over \partial \phi} + 
{\partial u_{\phi} \over \partial r} - {u_{\phi}\over r } \right) \\ 
Q_{\phi\phi} &=& 2 \mu_{\rm v} \left( {1\over r} {\partial u_{\phi} \over \partial \phi} + 
{u_{r} \over r} - {1\over 3} (\bl \nabla \cdot \bl u)  \right) 
\end{eqnarray} 
 
 
\section{Tests and accuracy} 
 
In this section we provide the results of two test problems probing the 
advection scheme and, especially, the conservation of the specific angular momentum. 
The results of other tests will be given elsewhere. 
 
\subsection{Test 1: relaxation problem} 
 
The van Leer advection scheme was tested on a ''relaxation'' problem, 
in which a disk of constant density and velocity field proportional to $r$ ($u_{r}=u_{0} r$) 
is set, and the density is allowed to decrease. 
For this problem, the analytic solution of the continuity equation is $\Sigma(t)=\Sigma_0 
e^{-2 u_0 t}$ and  it gives $\Sigma=6.14 \times 10^{-6}\, 
\Sigma_0$ at $u_0 t=6$. The numerical solution for this problem gives density values 
of $5.98\times 10^{-6}\, \Sigma_0$ for the resolution of 256 radial grid 
points. When the density has decreased by nearly six orders of magnitude, 
its radial profile remains perfectly flat and the relative error is only $2.6\%$. 
 
\subsection{Test 2: Conservation of specific angular momentum} 
 
An important point of concern when numerically studying the dynamics of 
galaxies is the ability of a code to conserve specific angular momentum. 
A comprehensive test problem on specific angular momentum conservation 
that covers advection as well as pressure 
and gravity terms in the momentum equations was designed by Norman et al. 
(\cite{norman80}).  These authors considered the isothermal gravitational collapse  of 
rotating axisymmetric prestellar clouds in cylindrical geometry. 
For a fluid with no 
mechanism for redistributing angular momentum, the mass $M(K)$ in the cloud 
with specific angular momentum $K=\Omega r^2$ less than or equal to $K$ 
is a constant of motion. Deviations from the initial spectrum $M(K)$ 
reveal a redistribution of angular momentum. 
We consider 
a similar test problem of isothermal gravitational collapse of a flattened 
axisymmetric prestellar cloud (further referred as a ``disk'') 
that can be studied in the thin-disk approximation (see 
e.g. Basu \cite{basu97}). 
This problem should test the ability of our polytropic model (for which 
we set $\gamma=1$ to switch to the isothermal regime) to conserve 
the specific angular momentum. For a uniformly rotating gas disk 
($\Omega=\Omega_0$) with the radial surface density distribution given by Basu 
(\cite{basu97}), 
\begin{equation} 
\Sigma(r)={r_0 \Sigma_0 \over \sqrt{r^2+r_0^2}}, 
\end{equation} 
the initial specific angular momentum spectrum is   
\begin{equation} 
M(K)=2 \pi \Sigma_0 r_0^2 \left( \sqrt{1+ {K\over K_0}} -1 \right), 
\end{equation} 
where $K_0=\Omega_0 r_0^2$, $\Sigma_0$ is the central surface density, and 
$r_0=c_{\rm s}^2/(1.5 G \Sigma_0)$ is the characteristic scale length which 
is comparable to the Jeans length $\lambda_{\rm J}=c_{\rm s}^2/(G\Sigma)$ 
of a nonrotating cloud. We choose $\Sigma_0=3.55\times 10^{-2}$~gm~cm$^{-2}$, 
$c_{\rm s}=0.188$~km~s$^{-1}$ (equivalent to $T=10$~K), $\Omega_0=0.4$~km~s$^{-1}$~pc$^{-1}$, 
and the mean molecular weight 
2.33 (molecular hydrogen with a $10\%$ admixture of atomic helium). 
The disk has the inner and outer radii of $r_{\rm in}=10$~AU and $r_{\rm 
out}=20000$~AU. We introduce a ``sink cell'' at $r<10$~AU and impose a free inflow 
inner boundary condition and a constant mass and volume outer boundary 
condition.  The mass of our disk is $2.45~M_\odot$. 
 
The ratios of rotational and thermal energies to the gravitational 
energy of the initial configuration are $0.7\%$ and $25\%$, respectively. 
The disk is thus gravitationally unstable and is allowed to collapse under 
its own gravity. The initial theoretical spectrum $M(K)$ 
is shown in Fig.~\ref{fig13} by the solid line. 
The early evolution is characterised by a slow gravitational contraction and a 
subsequent runaway collapse. When the density in the sink cell exceeds 
13.3 g~cm$^{-2}$, the central protostar is assumed to form (due 
to a transition to the adiabatic phase). The filled circles plot $M(K)$ computed at 
$t=4300$~yr  after the formation of the central protostar (0.185~Myr after the beginning of the simulation). 
At this time, approximately $0.3~M_\odot$ has been accreted by the 
protostar. 
As is clearly seen, $M(K)$ merges with the initial spectrum (except for 
the end points where a minimal deviation is observed), indicating virtually 
no angular momentum redistribution due to either physical or numerical reasons. 
This is expected in the axisymmetric collapse with very little 
numerical diffusion. 
 
\begin{figure} 
    \resizebox{\hsize}{!}{\includegraphics{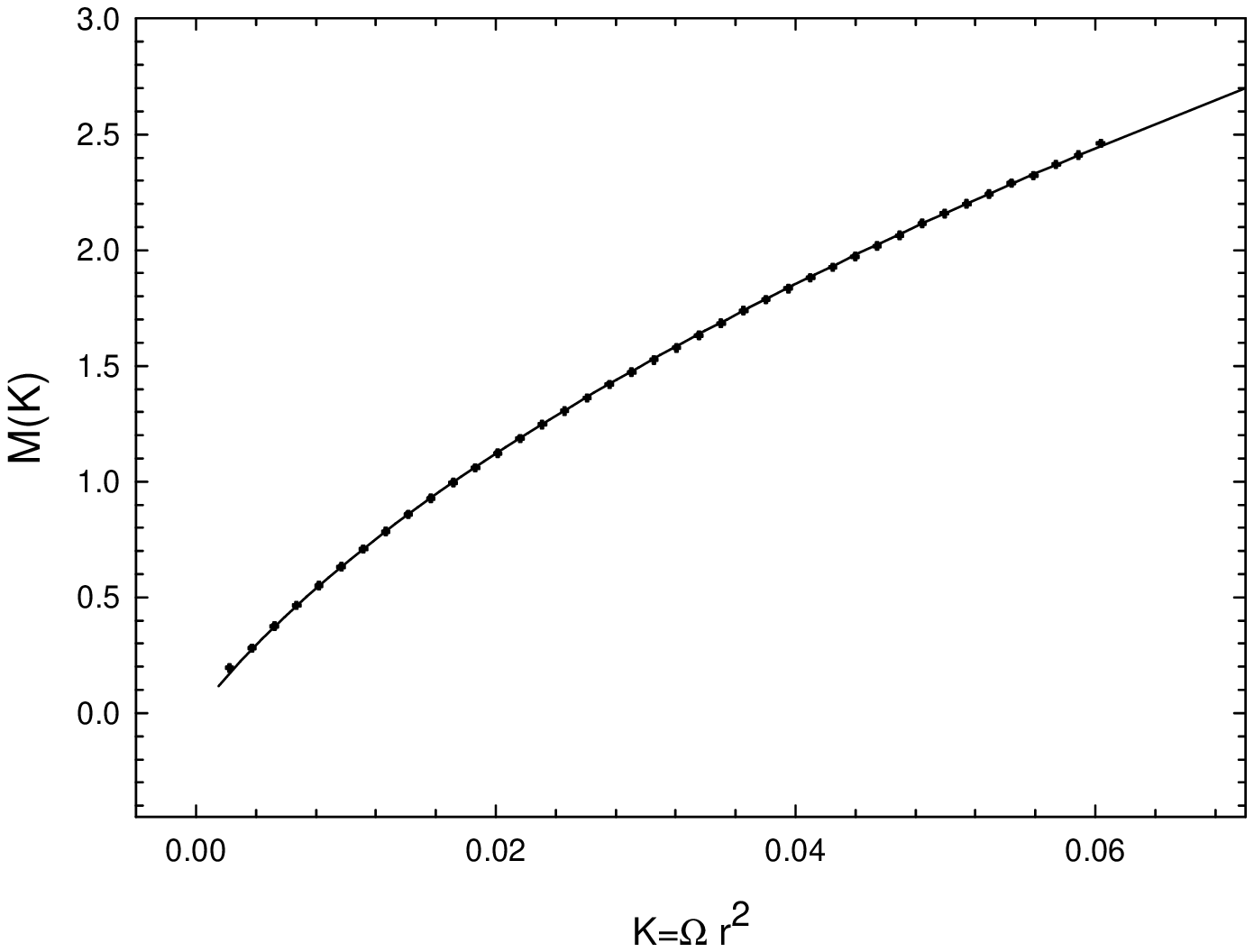}} 
      \caption{The specific angular momentum spectrum of a collapsing flattened 
      cloud. The quantity $M(K)$ is the total mass in the cloud with specific 
      angular momentum less than or equal to $K=\Omega r^2$. The solid 
      line shows the theoretical spectrum at $t=0$~yr. The filled circles 
      gives $M(K)$ at 0.185~Myr after the beginning of the simulation when 
      the central protostar has already formed. A good agreement between 
      both spectra indicates virtually no angular momentum redistribution, 
      as is indeed expected for the axisymmteric collapse. 
} 
         \label{fig13} 
\end{figure} 
 

\begin{thebibliography}{} 
 
\bibitem[1991]{amendt91} 
     Amendt, P., Cuddeford, P., 1991, ApJ, 368, 79 
 
\bibitem[1979]{aoki79} 
     Aoki, S., Noguchi, M., Iye, M., 1979, PASJ, 31, 737 
 
\bibitem[1997]{basu97} 
     Basu, S., 1997, ApJ, 485, 240   
 
\bibitem[1989]{bertin89} 
     Bertin, G., Lin, C.C., Lowe S.A., Thurstans, R.P., 1989, ApJ, 338, 104 
 
   
\bibitem[1987]{binney87} 
     Binney, J., Tremaine, S., 1987, Galactic Dynamics, 
                Princeton Univ.\ Press 
 
\bibitem[1998]{binney98} 
     Binney, J., Merrifield, M. S., 1998, Galactic Astronomy, 
                Princeton Univ.\ Press 
 
%
 
\bibitem[1991]{cuddeford91} 
     Cuddeford, P., Amendt, P., 1991, MNRAS, 253, 427 
 
\bibitem[1998]{Evans} 
     Evans, N. W., Read, J. C. A., 1998, MNRAS, 300, 106 
 
\bibitem[1970]{freeman} 
     Freeman, K. C., 1970, ApJ, 160, 811 
 
\bibitem[1997]{Gerssen} 
     Gerssen J., Kuijken K., Merrifield M. P., 1997, MNRAS, 288, 618   
 
\bibitem[1994]{giersz94} 
     Giersz, M., Spurzem, R., 1994, MNRAS, 269, 241 
 
\bibitem[1966]{JT} 
     Julian, W. H., Toomre, A., 1966, ApJ, 146, 810 
 
 
\bibitem[1994]{KT} 
     Kuijken, K., Tremaine, S., 1994  ApJ, 421, 178 
 
 
\bibitem[2000]{korchagin00} 
     Korchagin, V.I., Kikuchi, N., Miyama, S.M., et al., 2000,   
       ApJ, 541, 565 
 
\bibitem[2005]{Jalali} 
       Jalali, M. A., Hunter, C., 2005, ApJ, 630, 804 
 
\bibitem[1970]{larson70} 
     Larson, R.B., 1970, MNRAS, 147, 323 
 
\bibitem[1997]{laughlin97} 
     Laughlin, G., Korchagin, V., Adams, F.C., 1997, ApJ, 477, 410 
 
\bibitem[1998]{laughlin98} 
     Laughlin, G., Korchagin, V., Adams, F.C., 1998, ApJ, 504, 945 
 
\bibitem[1990]{louis90} 
     Louis, P., 1990, MNRAS, 244, 478 
 
 
\bibitem[1980]{norman80} 
     Norman, M. L., Wilson, J. R., Barton, R. T., 1980, ApJ, 239, 968 
 
\bibitem[2002]{orlova02} 
     Orlova, N., Korchagin, V.I., Theis, Ch., 2002, A\&A, 384, 872 
 
\bibitem[1997]{Pichon} 
     Pichon, C., Cannon, R. C., 1997, MNRAS, 291, 616 
 
 
\bibitem[1997]{polyachenko97b} 
     Polyachenko, V. L., Polyachenko, E. V., Strel'nikov, A. V., 1997, 
      Astron.\ Zhurnal, 23, 598 (translated Astron.\ Lett.\ 23, 525) 
 
\bibitem[1997]{samland97} 
     Samland, M., Hensler, G., Theis, Ch., 1997, ApJ, 476, 544 
 
\bibitem[1992]{stone92} 
     Stone, J. M., Norman, M. L., 1992, ApJS, 80, 753 
 
\bibitem[1992]{theis92} 
     Theis, Ch., Burkert, A., Hensler, G., 1992, A\&A, 265, 465 
 
\bibitem[1964]{toomre64} 
     Toomre, A., 1964, ApJ, 139, 1217 
      
\bibitem[1981]{toomre81} 
     Toomre, A., in Fall S. M., Lynden-Bell D., eds., The Structure 
     and Evolution of Normal Galaxies. Cambridge Univ. Press, Cambridge 
 
 
\bibitem[2006]{VB} 
     Vorobyov, E. I., Basu, S., 2006, ApJ, in press
 
 
 
\bibitem[1976]{Zang} 
     Zang, t. A., 1976, PhD thesis, Massachusetts Institute of Technology 
 
\end{thebibliography}
\end{document}